\documentclass[lettersize,journal]{IEEEtran}
\usepackage{amsmath,amsfonts}
\usepackage{makecell}
\usepackage{algorithmic}
\usepackage{algorithm}
\usepackage{array}
\usepackage{overpic}
\usepackage{textcomp}
\usepackage{stfloats}
\usepackage{url}
\usepackage{verbatim}
\usepackage{graphicx}
\usepackage{cite}
\usepackage{subcaption}
\usepackage{epstopdf}
\usepackage{autobreak}
\usepackage{multirow}
\usepackage[absolute,overlay]{textpos}
\usepackage{booktabs}
\usepackage{soul, color, xcolor}
\usepackage{enumitem}
\usepackage{enumitem}
\usepackage{textcomp}
\usepackage{upgreek}
\usepackage[draft]{hyperref}

\begin{document}

\title{A Simultaneous Clustering and Tracking Algorithm for Capturing Cluster-Level Spatial Consistency in 6G Wireless Channels}

\author{Jiaxin~Lin, Pan~Tang, Jianhua~Zhang, Zhaowei~Chang, Peijie~Liu, Yufeng~Qin, Ke~Chen, Huixin~Xu and Byonghyo~Shim

\thanks{This work was supported in part by the National Key R\&D Program of China under Grant 2023YFB2904805, in part by the National Natural Science Foundation of China under Grant 62201086, in part by the Beijing Natural Science Foundation under Grant L243002, and in part by the Beijing University of Posts and Telecommunications-China Mobile Research Institute Joint Innovation Center. \textit{(Corresponding author: Pan Tang)}}

\thanks{
Jiaxin~Lin, Pan~Tang, Jianhua~Zhang, Zhaowei~Chang, Peijie~Liu, Yufeng~Qin, Ke~Chen and Huixin~Xu are with the State Key Lab of Networking and Switching Technology, Beijing University of Posts and Telecommunications, Beijing, 100876, P.R. China 
(email: linjx@bupt.edu.cn; tangpan27@bupt.edu.cn; jhzhang@bupt.edu.cn; changzw12345@bupt.edu.cn; Liupj@bupt.edu.cn; qinyf@bupt.edu.cn; chen-ke@bupt.edu.cn; xuhuixin@bupt.edu.cn) 

Byonghyo~Shim is with the Institute of New Media and Communications, Department of Electrical and Computer Engineering, Seoul National University, Seoul, 08826, Republic of Korea (e-mail: bshim@snu.ac.kr)}}

\markboth{Journal of \LaTeX\ Class Files,~Vol.~14, No.~8, August~2021}%
{Shell \MakeLowercase{\textit{et al.}}: A Sample Article Using IEEEtran.cls for IEEE Journals}

\IEEEpubid{0000--0000/00\$00.00~\copyright~2021 IEEE}

\maketitle
\begin{abstract}
Spatial consistency is a fundamental physical property of wireless channels that reflects the smooth evolution of the channel between spatial locations. At the cluster level, it requires similar multipath components (MPCs) remain grouped into the same clusters as the transceivers move, enabling consistent cluster tracking. Cluster-level spatial consistency is essential for realistic cluster-based channel models, especially for potential 6G techniques such as massive MIMO, integrated sensing and communication, and terahertz (THz) communication. However, existing clustering and tracking methods do not fully exploit spatial correlations of MPCs. In tracking-after-clustering, clustering and tracking are decoupled, while joint clustering-and-tracking mainly relies on cluster centers from the previous snapshot. In this work, we propose a Mahalanobis-distance-based simultaneous clustering and tracking (MD-SCT) algorithm to capture the joint distribution of clustered MPCs in the delay, angular and spatial domains. Under Mahalanobis distance, MPCs in successive snapshots are associated with existing clusters, thereby inherently tracking while clustering. The algorithm is further applied in the sub-THz band. Performance is evaluated using mean square successive difference and gradient change rate. The results demonstrate that the proposed algorithm yields smoother cluster evolution. This improves the reliability of clustered channels for spatial consistency modeling in 6G.
\end{abstract}
	
	
\section{Introduction}\label{section1}
\IEEEPARstart{C}{hannel} is the medium in which the radio wave propagates, linking the transmitter (TX) and receiver (RX), and its properties affect the performance of the wireless communication system \cite{Molisch}. Channel modeling serves as a basis for the design and evaluation of 6G wireless systems \cite{A-3-Surv, BShim6G}. Spatial consistency refers to the similarity of channel propagation effects experienced by transceivers in similar spatial locations \cite{1-1-1,1-1-19}. At the cluster level, multipath components (MPCs) with similar delay and angular are typically grouped into clusters \cite{MHYJSACFR3, KPMTP}. Channel measurement campaigns have provided empirical evidence of cluster-level spatial consistency \cite{1-1-3,1-1-6}. Cluster-based channel modeling allows for separate consideration of inter- and intra-cluster characteristics, thereby reducing the number of statistical parameters \cite{1-2-7,1-2-12}. For this reason, cluster-based channel model has been widely adopted in both academia and standardization \cite{1-2-13}, including the 3rd Generation Partnership Project (3GPP) spatial channel model \cite{3GPPSCM}, Wireless World Initiative New Radio (WINNER) \uppercase\expandafter{\romannumeral2} \cite{WINNER2} channel model and the European Cooperation in Science and Technology (COST) 2100 channel model \cite{COST}, among others. The 5G standard channel model, International Telecommunication Union Radiocommunication Sector (ITU-R) M.2412, has taken spatial consistency as one of its advanced modeling components to ensure that cluster-specific and ray-specific random variables evolve smoothly \cite{2412}. Spatial consistency is also being considered as a key concept for upcoming 6G technologies. For massive MIMO and integrated sensing and communication, spatial consistency ensures stable multipath structures that are essential for accurate beam tracking and environment-aware sensing \cite{1-1-11,MHYJSACFF,MIMOWH1,ISACJSAC}. In the terahertz (THz) band, the high sensitivity of narrow beams to misalignment \cite{BShim-2, XHX, 1-1-7,1-1-12} underscores accurate channel modeling with spatial consistency.

\IEEEpubidadjcol
    
Clustering and tracking methods extract cluster-level spatial consistency from channel data and support channel analysis and modeling. While early studies used visual inspection, the rise of data-driven approaches has motivated the development of automated methods \cite{CSI,DTC}. Two main approaches namely tracking-after-clustering and joint clustering-and-tracking have been proposed \cite{1-2-13}. In the tracking-after-clustering approach, clustering is performed independently on each snapshot using algorithms such as KPowerMeans \cite{1-2-7-13} and density-based spatial clustering of applications with noise (DBSCAN) \cite{1-2-6-2}, typically based on the MPC distance---a Euclidean distance computed over normalized MPC parameters \cite{1-1-3, KPMTP}. 
Cluster tracking is then performed by associating clusters in adjacent snapshots. One representative method compares cluster centroids in adjacent snapshots by using the multipath component distance (MCD) \cite{1-2-13-10}. Other studies incorporated additional tracking features, including cluster centroid, shape, and density \cite{TAPCommNew}, and scattering-point representations \cite{R1-4}. In addition, features extracted from the power-angle spectrum (PAS), such as size, mass center, and shape, were used in \cite{R1-7}. For tracking-based MPC clustering, \cite{R1-5} constructs moving probabilities for MPC associations and then determines clusters from the resulting probabilities. In this two-step structure, tracking depends on the clustering results obtained in each snapshot. For the joint clustering-and-tracking approach, information from previous snapshots is incorporated into current clustering and tracking. In \cite{1-2-13-13}, the current snapshot is clustered with reference to the clustering result of the previous snapshot. In \cite{1-2-13-12}, a Kalman filter predicts cluster centroids that are used to initialize KPowerMeans in the current snapshot, whereas \cite{R1-8} uses a matching procedure to incorporate Kalman predictions before KPowerMeans is applied to the remaining MPCs. In existing joint clustering-and-tracking methods, information from previous snapshots is introduced mainly through cluster centroids or their predictions. Regarding evaluation metrics, clustering and tracking are usually considered separately. Clustering indices such as Davies–Bouldin (DB) \cite{DB} and Calinski–Harabasz (CH) indices \cite{CH} are defined for individual snapshots, while cluster tracking can be evaluated by the gradient change rate (GCR) \cite{TAPCommNew}, which characterizes cluster evolution across snapshots. As a result, the existing metrics provide only a partial assessment and cannot jointly reflect clustering quality at each snapshot and tracking consistency across snapshots.

For cluster-level spatial consistency, the evolution of clustered MPCs across snapshots is jointly reflected in delay, angle, and transceiver location. Therefore, the considered problem is different from general clustering tasks, because the objective is not only to obtain reasonable clustering results in individual snapshots, but also to maintain consistent cluster tracking and continuous cluster-parameter evolution across successive snapshots. Existing methods mainly improve cluster association by using local matching, representative cluster features, or information from previous snapshots. In this work, we propose a Mahalanobis-distance-based simultaneous clustering and tracking (MD-SCT) algorithm, which assigns newly observed MPCs according to their consistency with the parameter distribution of existing clusters. The MD-SCT algorithm exploits the covariance structure of delay, angle, and transceiver position formed by previously assigned MPCs, thereby improving cluster-level spatial consistency. The main contributions of this paper are summarized as follows:
		\begin{enumerate}
			\item We propose the MD-SCT algorithm to improve cluster-level spatial consistency by exploiting the distribution structure of delay, angle, and transceiver position formed by previously assigned MPCs.
			\item We apply the MD-SCT algorithm to sub-THz channels at 132 GHz generated via measurement-based deterministic ray tracing in an industrial IoT scenario.
			\item We introduce the mean square successive difference (MSSD) as a novel metric to quantify the spatial consistency of clustering statistics across successive snapshots. 
	\end{enumerate}
	
	\section{Algorithm Based on the Mahalanobis Distance}\label{section2}
	In this section, we describe the MD-SCT algorithm. Also, we briefly explain corresponding clustering and tracking validation indices.
	\subsection{Proposed Clustering and Tracking Algorithm}
	The MD-SCT algorithm (as shown in Algorithm \ref{alg1}) achieves simultaneous clustering and tracking of MPCs. The Mahalanobis distance incorporates the covariance matrix of MPC parameters, exploiting the correlations among delay, angle, and location parameters. It measures the distance of an observed MPC from the parameter distribution of an existing cluster. By incorporating parameters from previously classified snapshots into the cluster distribution, associating new MPCs with existing clusters inherently achieves tracking while clustering.
    
    In detail, initial clusters in small spatial regions are first obtained by applying conventional clustering methods, such as the partition-based KPowerMeans \cite{1-2-7-13} and the density-based DBSCAN \cite{1-2-6-2}, to densely sampled MPCs. The high similarity of dense MPCs ensures effective conventional clustering and accurate covariance matrices.

	\begin{algorithm}[t]
		\caption{The Mahalanobis distance based MPCs clustering algorithm for spatial consistency}\label{alg:alg1}
		\renewcommand{\algorithmicrequire}{\textbf{Input:}}
		\renewcommand{\algorithmicensure}{\textbf{Output:}}
		\begin{algorithmic}[1]
			\REQUIRE A set of MPC vector, $\Omega=\{\mathbf{x}_1, \mathbf{x}_2, \ldots, \mathbf{x}_m,$ $\mathbf{x}_{m+1}, \ldots, \mathbf{x}_n\}$, where $\mathbf{x}$ consists of multidimensional parameters such as delay ($\tau$), azimuth angle of departure ($\phi_{\text{AOD}}$), azimuth angle of arrival angle ($\phi_{\text{AOA}}$), zenith angle of departure ($\theta_{\text{ZOD}}$), zenith angle of arrival ($\theta_{\text{ZOA}}$), power ($p$), and location ($x_{i,\mathrm{TX}}^{\mathrm{POS}}, y_{i,\mathrm{TX}}^{\mathrm{POS}}, z_{i,\mathrm{TX}}^{\mathrm{POS}},x_{i,\mathrm{RX}}^{\mathrm{POS}}, y_{i,\mathrm{RX}}^{\mathrm{POS}}, z_{i.\mathrm{RX}}^{\mathrm{POS}}$). \\
			\ENSURE The cluster index $c_i$ for MPCs $\mathbf{x}_i$. \\
			\STATE Cluster subset $\{\mathbf{x}_1, \mathbf{x}_2, \ldots, \mathbf{x}_m\}$ of $\Omega$ using conventional clustering methods where the locations are dense and similar. Then, the subset are divide into $K_{1}$ clusters, each MPC $\mathbf{x}_i$ is assigned a cluster index $c_i$ within the range $[1,K_{1}]$.
			\FOR{$i = m+1  \to \,  n$}
			\FOR{$k = 1  \to \,  K_{1}$}
			\STATE Calculate the $D_{\text{Mah},i,k}$ between $\mathbf{x}_i$ and the set of MPCs with cluster $k$ by (1), where $\mathbf{x}_i$ is composed of $\tau_{i}, \phi_{i,\text{AOD}}, \phi_{i,\text{AOA}}, \theta_{i,\text{ZOD}},\theta_{i,\text{ZOA}},$ locations.
			\ENDFOR
			\STATE Obtain the $\tilde{k}$ that minimizes $D_{\text{Mah},i,k}$ of $\mathbf{x}_i$.
			\IF {$ D_{\text{Mah},i,\tilde{k}} <  D_{\text{th}} $}
			\STATE Mark the cluster index $c_i$ of $\mathbf{x}_i$ as $\tilde{k}$.
			\ELSE
			\STATE Mark the cluster index $c_i$ of $\mathbf{x}_i$ as outlier.
			\ENDIF
			\ENDFOR
			\STATE Cluster all MPCs marked as outliers into $K_{2}$ clusters labeled from $K_{1}+1$ to $K_{1}+K_{2}$ using conventional clustering methods.
			\RETURN {Cluster index $\mathbf{c}$}
		\end{algorithmic}	\label{alg1}
	\end{algorithm}
    
    The Mahalanobis distance $D_{\mathrm{Mah},i,k}$, which quantifies the distance between the MPC vector $\mathbf{x}_i$ and cluster $k$, can be expressed as
	\begin{equation}\label{equMahD}
		D_{\mathrm{Mah},i,k} =\sqrt{{\left ( \mathbf{x}_i  -\mathbf{y}_{i,k}  \right )} ^\mathrm{T}\cdot \mathbf{C}_{i,k} ^{-1} \cdot {\left ( \mathbf{x}_i  -\mathbf{y}_{i,k}  \right )}  }  ,  
	\end{equation}
	where the parameter vector $\mathbf{x}_i$ of the $i$-th MPC can consist of elements including delay ($\tau_{i}$), azimuth angle of departure (AOD) ($\phi_{i,\mathrm{AOD}}$), azimuth angle of arrival (AOA) ($\phi_{i,\mathrm{AOA}}$), zenith angle of departure (ZOD) ($\theta_{i,\mathrm{ZOD}}$) and zenith angle of arrival (ZOA) ($\theta_{i,\mathrm{ZOA}}$), location ($x_{i,\mathrm{TX}}^{\mathrm{POS}}, y_{i,\mathrm{TX}}^{\mathrm{POS}}, z_{i,\mathrm{TX}}^{\mathrm{POS}},x_{i,\mathrm{RX}}^{\mathrm{POS}}, y_{i,\mathrm{RX}}^{\mathrm{POS}}, z_{i.\mathrm{RX}}^{\mathrm{POS}}$). The vector $\mathbf{y}_{i,k}$ and the matrix $\mathbf{C}_{i,k}$ denote the mean vector and covariance matrix, respectively, of cluster $k$ for the $i$-th MPC. The mean vector $\mathbf{y}_{i,k}$ is given by:
	\begin{equation}\label{equCluDistr}
		\mathbf{y}_{i,k} = \frac{1}{N_{i,k}}\mathbf{1}_{N_{i,k}}^{\mathrm{T}} \mathbf{Y}_{i,k} = \frac{1}{N_{i,k}}\begin{bmatrix}1\:1\cdots\:1\end{bmatrix} \begin{bmatrix} \mathbf{x}_{k_1} \\ \mathbf{x}_{k_2} \\ \vdots \\ \mathbf{x}_{N_{i,k}} \end{bmatrix},
	\end{equation}
	where $N_{i,k}$ denotes the number of MPCs contained in the clustered distribution matrix $\mathbf{Y}_{i,k}$. The $(p,q)$ entry of the covariance matrix $\mathbf{C}_{i,k}$ is defined as
	\begin{equation}\label{equCov}
		C_{i,k}\left (p,q  \right ) =\frac{1}{N_{i,k}-1}  {\textstyle \sum_{j=1}^{N_{i,k}}} \left ( Y_{i,k}\left ( j,p \right ) -\mu _{p} \right ) \left ( Y_{i,k}\left ( j,q \right ) -\mu _{q} \right ) .
	\end{equation}
	where 
	\begin{equation}\label{equCovmu}
		\mu_{p}=\frac{1}{N_{i,k}}  {\textstyle \sum_{j=1}^{N_{i,k}}} Y_{i,k}\left ( j,p \right ),
	\end{equation}
	is the mean of the $p$-th parameter, and $\mu_{q}$ is defined similarly. $Y_{i,k}\left ( j,p \right )$ is the $(j,p)$-th entry of clustered distribution matrix $\mathbf{Y}_{i,k}$. 

     A sufficiently dense initialization region provides enough MPC samples to estimate the initial covariance matrices used in the Mahalanobis-distance-based assignment. Insufficient samples may yield a rank-deficient or ill-conditioned covariance matrix, affecting the subsequent assignment. Covariance regularization, such as diagonal loading $\mathbf{C}_{i,k}^{\mathrm{reg}}=\mathbf{C}_{i,k}+\epsilon\mathbf{I}$ with $\epsilon>0$, can be adopted to improve numerical stability.
     When the covariance matrix $\mathbf{C}_{i,k}$ is reduced to the identity matrix, the parameters are uncorrelated. The Mahalanobis distance incorporates correlations among the multidimensional parameters and the deviation of each MPC from the cluster mean, and it is inherently scale-invariant~\cite{2-1-5}. Since the covariance matrix is computed from all MPCs assigned to a cluster across previous snapshots, the proposed method naturally accumulates historical information beyond the most recent snapshot. If $D_{\text{Mah},i,\tilde{k}} < D_{\text{th}}$, MPC $i$ is assigned to cluster $\tilde{k}$ and retained for the next iteration. Otherwise, it is marked as an outlier. All marked outliers are then re-clustered by the same clustering method as in the initialization step (e.g., KPowerMeans or DBSCAN), and the resulting clusters are initialized as new clusters, corresponding to cluster birth. For cluster death, an existing cluster is regarded as inactive if no MPC is assigned to it over consecutive snapshots.

Compared with Euclidean-based distance calculation, the additional complexity of the Mahalanobis distance mainly comes from the inverse covariance matrix $\mathbf{C}_{i,k}^{-1}$. For an MPC vector with $N_{\mathrm{par}}$ parameters, such as delay, departure angles, and arrival angles, Euclidean-based distance calculation requires $O(N_{\mathrm{par}})$ complexity. For the Mahalanobis distance, directly recomputing the inverse of the updated covariance matrix after the cluster update would require $O(N_{\mathrm{par}}^3)$ complexity. This complexity can be reduced to $O(N_{\mathrm{par}}^2)$ by iteratively updating the inverse covariance matrix from the previous one and the newly assigned MPC using the Woodbury matrix inversion lemma \cite{Woodbury}.

	The proposed MD-SCT algorithm can be further extended to handle temporary interruptions in cluster observation through per-cluster local memory and dormant-state retention. Each cluster updates its mean vector and covariance matrix using only its most recently observed MPCs. When a cluster remains unobserved over a number of consecutive snapshots, it is retained as a dormant candidate instead of being immediately discarded. The memory-and-dormancy mechanism allows clusters to survive brief observation gaps, thereby improving robustness to temporary blockage or the short-term disappearance of MPCs along the RX route.
	
	\subsection{Clustering and Tracking Validation Indices}\label{section2-2}
	This section introduces two indices: MSSD and GCR to evaluate the clustering and tracking. MSSD measures the consistency of clustering statistics sequences, while GCR focuses on the evolution of tracked cluster-center parameter trajectories.
	\subsubsection{Mean square successive difference (MSSD)}
	MSSD quantifies short-term variability by measuring the average squared difference between adjacent data in a sequence. The MSSD of a sequence $\{x_1, x_2, \dots, x_n\}$ is defined as follows. Note that lower MSSD value indicates smaller fluctuation and higher consistency across the sequence.
	\begin{equation}
		s_{\text{MSSD}} = \frac{1}{n - 1} \sum_{i=1}^{n-1} (x_{i+1} - x_i)^2.
	\end{equation}
	
	The sequence of clustering statistics, e.g., DB index \cite{DB}, CH index \cite{CH}, and standard deviations of MPC parameters, is used as $\{x_i\}$. Lower MSSD values of clustering statistics indicate higher spatial consistency across successive snapshots. When invalid values appear in a sequence, only adjacent pairs with two valid values are included in the MSSD calculation. Note that the angular standard deviation of MPC parameters is computed using circular statistics \cite{AngSTDref}. The calculation for AOA standard deviation is given by
		\begin{equation}\label{equAngleSTD}
			\sigma_{\text{AOA},i,k} =\sqrt{2\times \left ( 1-\frac{\left | \sum e^{\text{j} \phi_{\text{AOA},i,k}}   \right | }{N_{i,k}}  \right ) },  
		\end{equation}
where $\mathrm{j}$ is the imaginary unit, $\phi_{\text{AOA},i,k}$ is in radians, $\left| \sum e^{\text{j} \phi_{\text{AOA},i,k}} \right | /{N_{i,k}}$ is the mean resultant vector length. The angular standard deviation is thus bounded. 

	\subsubsection{Gradient change rate (GCR)}\label{secGCR}
	The GCR quantifies the accuracy of cluster-tracking results in time-varying channels by characterizing the temporal and spatial evolution of dynamic clusters \cite{TAPCommNew}.   
    For result comparison, this work uses a modified GCR. It keeps the fitted cluster-center trajectory, but replaces the summation over discrete route samples with an integration over the spatial route variable $r$. The fitted cluster-center trajectory for cluster $k$ is written as
    
\begin{equation}\label{equNPoly}
	\mathbf{g}_{k}(r)=\sum_{m=0}^{n}\mathbf{w}_{m,k}r^{m},
\end{equation}
where $\mathbf{g}_{k}(r)$ denotes the fitted trajectory of the cluster-center parameters, $r$ denotes the spatial route variable, $n$ is the polynomial order, and $\mathbf{w}_{m,k}$ is the corresponding coefficient vector. The coefficients are obtained by minimizing the mean square error. The modified GCR metric used for result comparison is defined as
\begin{equation}\label{equGCR}
	s_{\mathrm{GCR}}=\frac{1}{K}\sum_{k=1}^{K}
\int_{r_{k,1}}^{r_{k,N_k}}
\left\|
\frac{d^{2}\mathbf{g}_{k}(r)}{dr^{2}}
\right\|_{1}
dr.
\end{equation}

Here, $K$ is the number of clusters, and $r_{k,1}$ and $r_{k,N_k}$ are the route-variable values at the start and end of the fitted trajectory, respectively. The $l_1$-norm sums the absolute second derivatives of the cluster-center parameter components. A lower $s_{\mathrm{GCR}}$ indicates smoother cluster-center evolution.




    \begin{figure*}[!htb]
    \subfloat[Measurement scenario photo]{\includegraphics[width=0.28\linewidth]{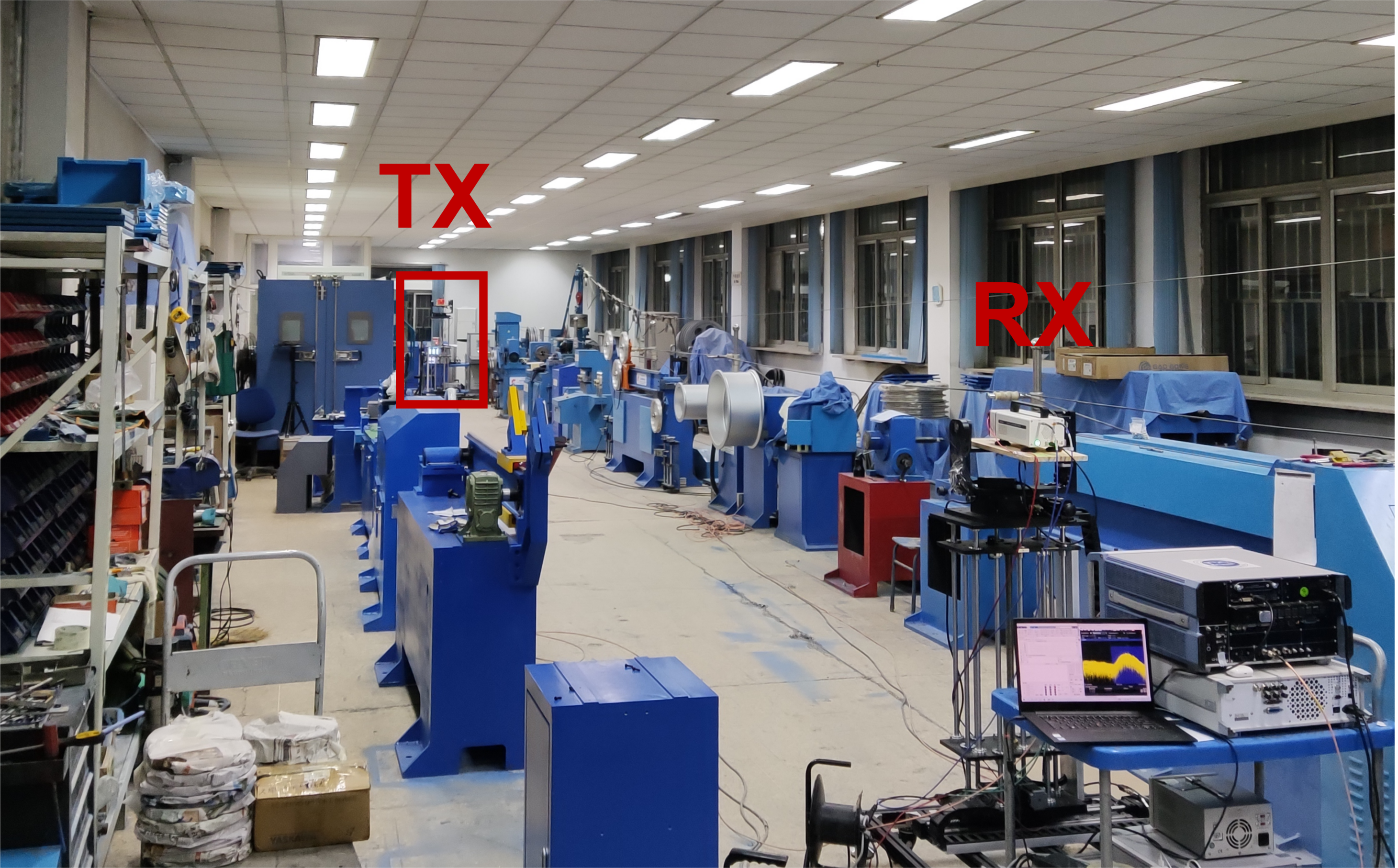}\label{FIG5a}}\hspace{2pt}
    \subfloat[Simulation layout and ray trajectories]{\includegraphics[width=0.70\linewidth]{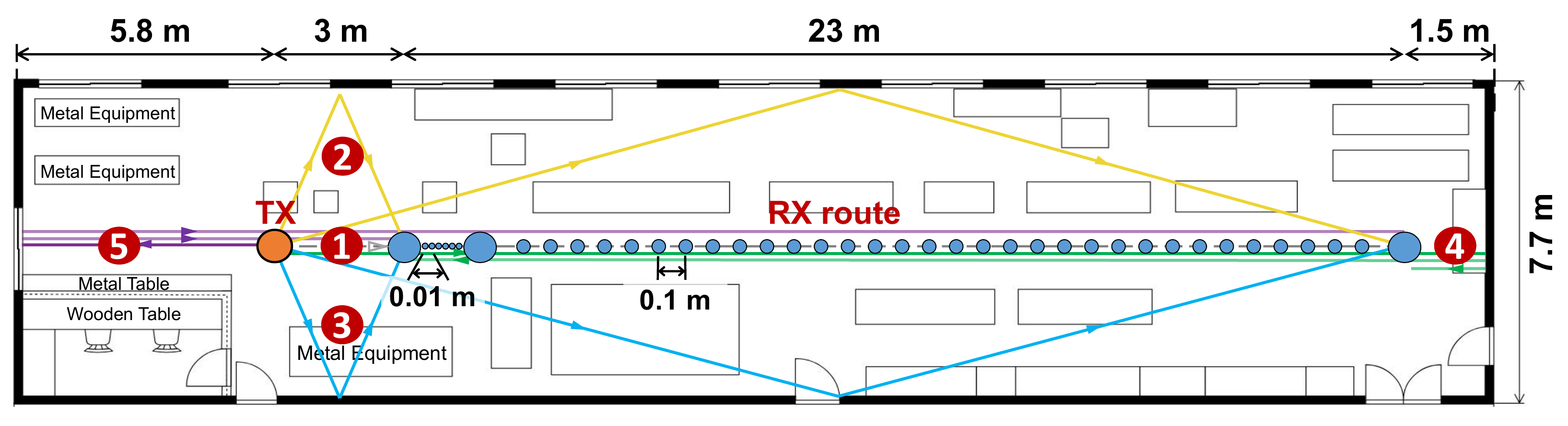}\label{FIG5b}}
    \vspace{-0.2cm}
    \caption{Measurement photo and simulation layout at 132 GHz for the industrial Internet of Things (IIoT) scenario. Dominant ray trajectories are marked in (b), including LOS (1) and first-order reflections (2--5).}
    \label{FIG5}
    \end{figure*}
    
	\section{Channel Data for Algorithm Validation}\label{section3}
	The proposed MD-SCT algorithm operates on extracted MPC parameters, including delay, angular parameters, and path power, making it applicable across 6G frequency bands when the required parameters are available. In this section, we present sub-THz channel data used for validation, generated via measurement-based deterministic ray tracing at 132 GHz in industrial Internet of Things (IIoT) scenario. The sub-THz band (100–300 GHz) offers abundant available spectrum resources \cite{A-3-Surv,1-1-12}. Its high directivity and spatial filtering from directional antennas or beamforming make communication sensitive to alignment \cite{1-1-7,XHX}. Therefore, accurate spatially consistent modeling is fundamental to reliable performance evaluation and system design for sub-THz communications.
    
    
    We conducte a double-directional measurement campaign at 132 GHz in IIoT scenario. The area is 33.3 m L $\times$ 7.7 m W $\times$ 3.0 m H, filled with metallic equipment and bounded by concrete walls. The TX is fixed at a height of 1.9 m and the RX at 1.4 m to represent a mobile user (Fig.~\ref{FIG5a}). Both antennas are swept to search MPCs across seven locations with 3–26 m separations. A sliding-correlation channel sounder generated a pseudorandom sequence that was upconverted to 132 GHz with 1.2 GHz bandwidth. More details can be found in \cite{LPJ}. For measurement-based simulation, the environment is modeled by the scenario layout. Then, the material electromagnetic properties are calibrated so that the simulated dominant MPC parameters match measurements, yielding a scalable high-angular-resolution dataset calibrated to field data.

\begin{figure}[!htb]
	\centering
	\setlength{\belowcaptionskip}{-0.3cm}
	\includegraphics[width=0.5\textwidth]{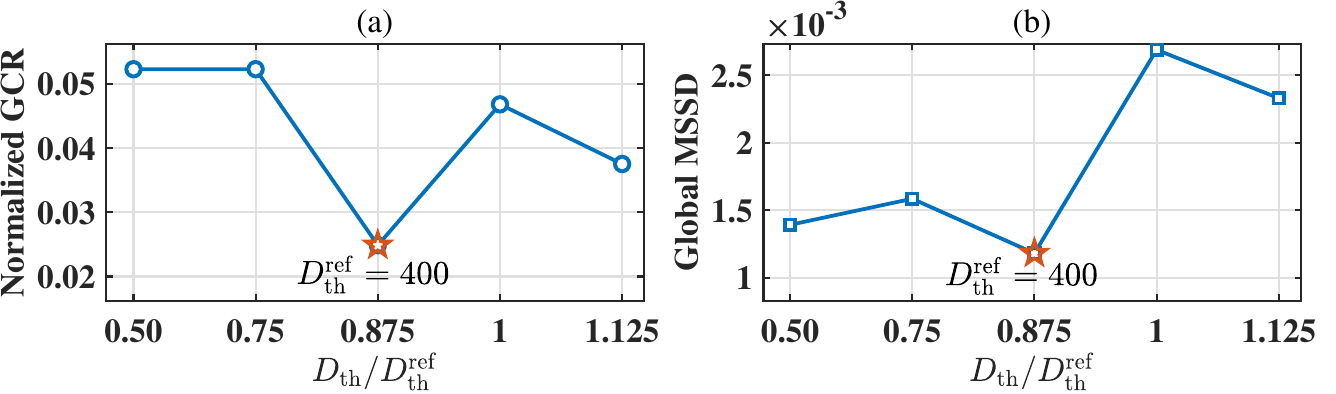}
	\caption{Sensitivity analysis of $D_{\mathrm{th}}/D_{\mathrm{th}}^{\mathrm{ref}}$ in terms of (a) normalized global gradient change rate (GCR) and (b) global mean squared successive difference (MSSD), where $D_{\mathrm{th}}^{\mathrm{ref}}=400$.}
	\label{fig:dth_sensitivity}
   \end{figure}

    As shown in Fig.~\ref{FIG5b}, the MPCs are then clustered and tracked along the RX route. The initial clustering subset consists of 100 snapshots from 3.00 to 3.99 m with a spacing of 0.01 m. A smaller RX snapshot spacing provides denser observations of the channel evolution and facilitates more reliable cluster association between adjacent snapshots, which is particularly important under high mobile speed.  Clustering is first performed using KPowerMeans \cite{1-2-7-13,KPMTP}, which updates the cluster centers with normalized-power weighting. KPowerMeans is first adopted to cluster MPCs in individual snapshots, because its clustering performance has been evaluated by measurement at THz band, with the average Silhouette coefficient \cite{1-2-8}. To determine the initial number of the cluster $K_1$, the cluster numbers are evaluated using the DB and CH indices, which yield $K_1=10$ as the optimal result. For the subsequent algorithm implementation, the MPCs are generated for TX--RX horizontal distances from 4 to 26 m at intervals of 0.1 m. Due to the affine-invariant property of the Mahalanobis distance, the only parameter that needs to be specified is the threshold $D_{\text{th}}$. For the considered 132 GHz IIoT case, $D_{\text{th}}$ is empirically set to 400, which gives the lowest GCR and MSSD in the threshold-sensitivity analysis as shown in Fig.~\ref{fig:dth_sensitivity}. 

    \begin{figure*}[!htbp]
		\vspace{-0.1cm}
		\centering
        \subfloat[Unclustered MPCs with five paths marked
]{\includegraphics[width=.35\linewidth]{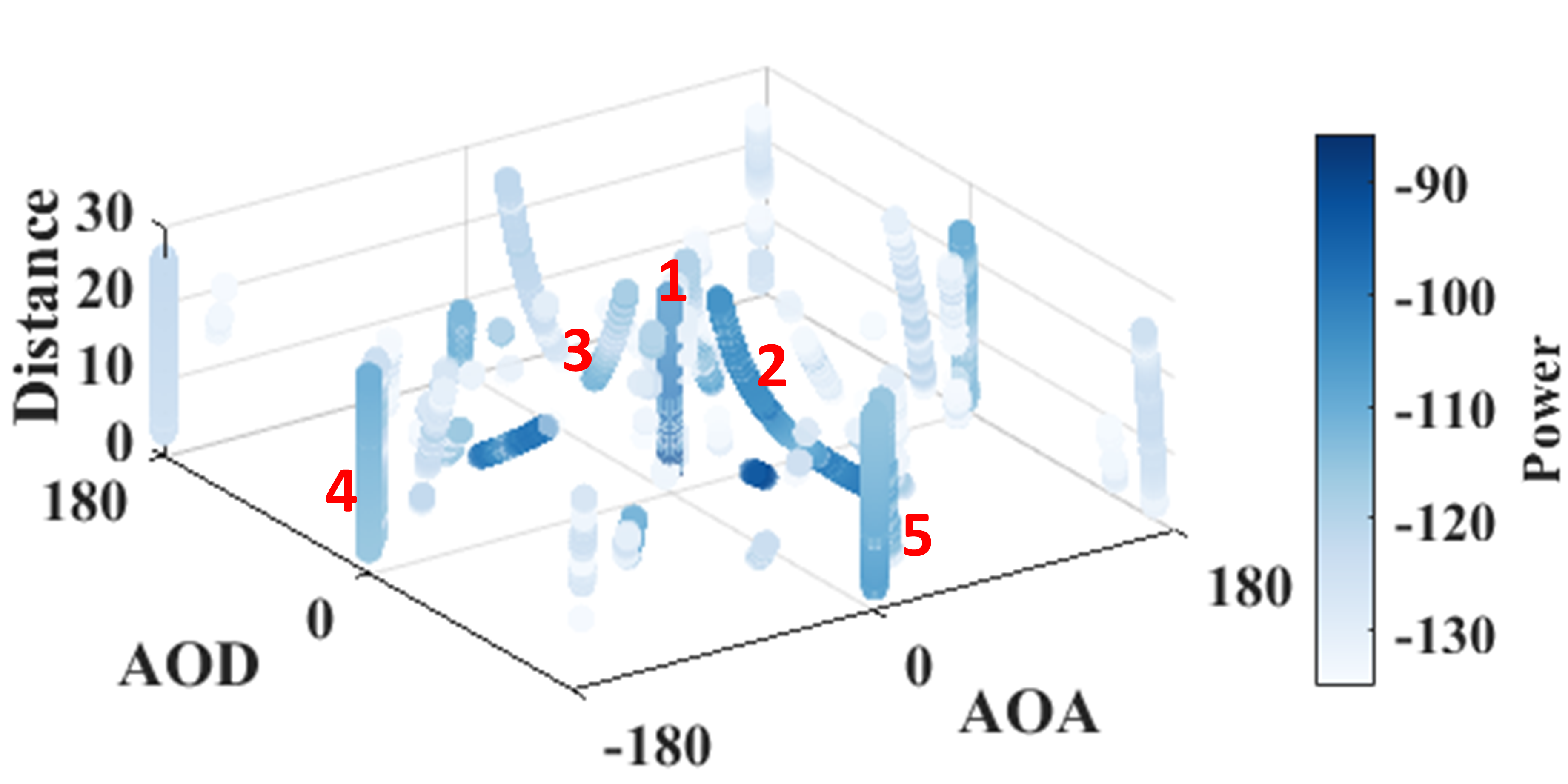}\label{fig3:01}}\hspace{3pt}
		\subfloat[MFM algorithm \cite{TAPCommNew}]{\includegraphics[width=.30\linewidth]{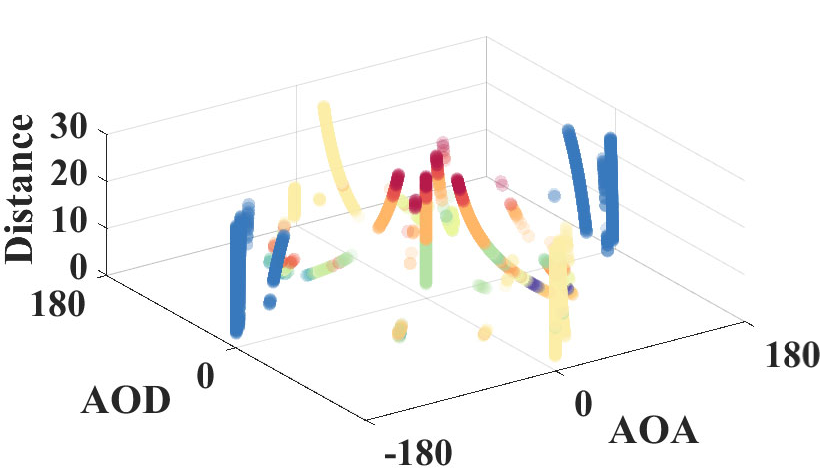}\label{fig3:1}}\hspace{3pt}
		\subfloat[MD-SCT algorithm]{\includegraphics[width=.30\linewidth]{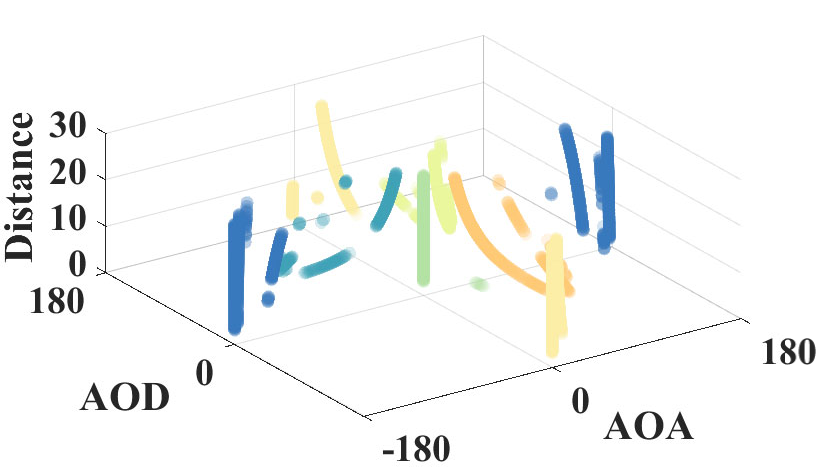}\label{fig3:2}}
		\caption{AOA-AOD of MPCs over TX-RX distance at 132 GHz. Five paths are marked in (a) (1: LOS; 2–5: first-order reflections). In (b)-(c), clusters are distinguished by color. In (b), paths 1–3 are split into discontinuous clustered segments, while} in (c), paths maintain continuous and distinguishable cluster labels.
		\label{fig33}
	\end{figure*}  

    \begin{table*}[!t]
	\setlength{\abovecaptionskip}{0cm} 
	\setlength{\belowcaptionskip}{0cm}
	\caption{Comparison of clustering and tracking metrics for evaluating cluster-level spatial consistency}
	\begin{center}
		\renewcommand\arraystretch{1.5}
		\tabcolsep=0.105cm
		\resizebox{\linewidth}{!}{
		\begin{tabular}{c|c|c|c|c|c|c|c|c|c|c|c|c}
			\toprule
			\hline
			\multirow{2}*{\raisebox{-1.0em}{Method}} 
			& \multirow{2}*{\raisebox{-1.4em}{\makecell[c]{Number of\\clusters}}}
			& \multirow{2}*{\raisebox{-1.0em}{Avg. length [m]}}
			& \multirow{2}*{\raisebox{-1.0em}{Avg. DB}}
			& \multirow{2}*{\raisebox{-1.0em}{Avg. CH}}
			& \multirow{2}*{\raisebox{-1.0em}{GCR}}
			& \multicolumn{7}{c}{MSSD} \\
			\cline{7-13}
			& & & & & 
			& \makecell[c]{\renewcommand\arraystretch{0.9}\small $s_\text{DB}$}
			& \makecell[c]{\renewcommand\arraystretch{0.9}\small $s_\text{CH}$\\[-0.1em]\small $(\times 10^2)$}
			& \makecell[c]{\renewcommand\arraystretch{0.9}\small $s_{\sigma,{\text{AOD}}}$\\[-0.1em]\small $(\times 10^{-3})$}
			& \makecell[c]{\renewcommand\arraystretch{0.9}\small $s_{\sigma,{\text{AOA}}}$\\[-0.1em]\small $(\times 10^{-3})$}
			& \makecell[c]{\renewcommand\arraystretch{0.9}\small $s_{\sigma,{\text{ZOD}}}$\\[-0.1em]\small $(\times 10^{-4})$}
			& \makecell[c]{\renewcommand\arraystretch{0.9}\small $s_{\sigma,{\text{ZOA}}}$\\[-0.1em]\small $(\times 10^{-4})$}
			& \makecell[c]{\renewcommand\arraystretch{0.9}\small $s_{\sigma,{\tau}}$\\[-0.1em]\small [ns$^2$]} \\ 
			\hline
			MFM \cite{TAPCommNew} 
			& 104 & 1.457 & 0.402 & 134.825 & 0.561
			& 0.007 & 10.925 & 7.011 & 7.122 & 2.268 & 2.374 & 41.570 \\ 
			\hline   
			MD-SCT     
			& 17 & 11.894 & 0.369 & 146.072 & 0.028
			& 0.003 & 7.226 & 2.708 & 1.715 & 1.669 & 1.660 & 25.994 \\ 
			\hline
			\bottomrule
		\end{tabular}
		}
	\end{center}
	\label{tab:mfm_mdsct_comparison}
	\vspace{-2em}
\end{table*}

	\section{Performance Validation}\label{section4}
	In this section, we perform the clustering and tracking on the channel in Section III. For comparison, we also test the multidimensional feature metric-based (MFM) cluster-tracking method \cite{TAPCommNew}. In the MFM method, clusters are first obtained in individual snapshots. Centroid, shape, and density features are then extracted from the obtained clusters, and cluster trajectories are identified according to the procedure in \cite{TAPCommNew}. The performance of the Mahalanobis-distance-based simultaneous clustering and tracking (MD-SCT) algorithm is then evaluated against the MFM method.
    

    The MPC clustering and tracking results are shown in Fig.~\ref{fig33}. As marked in Fig.~\ref{fig3:01}, five high-power dominant paths are identified, corresponding to the multipath components illustrated in Fig.~1b, including the LOS component (path 1) and four first-order reflections (paths 2--5). In Fig.~\ref{fig3:1} and Fig.~\ref{fig3:2}, clusters are distinguished by color. In Fig.~\ref{fig3:1}, as the TX-RX distance increases, paths 1--3 become close in the AOA-AOD domain and are merged into common clusters. Meanwhile, the cluster assignments change along the RX route, resulting in short discontinuous segments. Table~\ref{tab:mfm_mdsct_comparison} provides quantitative results complementary to Fig.~\ref{fig33}. The MFM algorithm produces 104 tracked clusters with an average tracked-cluster length of 1.457 m. The MD-SCT algorithm produces longer and more continuous cluster segments with stable label assignments, corresponding to 17 clusters with an average tracked-cluster length of 11.894 m.


    The MFM algorithm independently clusters each snapshot using KPowerMeans, where DB and CH indices determine the optimal clustering result at each snapshot, providing compact and well-separated clusters within individual snapshots. As the TX-RX distance increases, these paths become close in the AOA-AOD domain and are assigned to common clusters. The subsequent Kuhn--Munkres (KM)-based feature matching establishes one-to-one correspondence between clusters in adjacent snapshots. When the number or composition of clusters differs between adjacent snapshots, trajectories are terminated and new labels are assigned. The MD-SCT algorithm associates newly observed MPCs with existing clusters using the Mahalanobis distance computed from the historical cluster distribution, performing clustering and tracking simultaneously rather than as two separate stages. The average DB and CH values of MFM are 0.402 and 134.825, while those of MD-SCT are 0.369 and 146.072. Although the MFM algorithm optimizes DB and CH independently at each snapshot, MD-SCT achieves comparable cluster compactness and separation while producing fewer and longer cluster trajectories. On this route, the average Step-11 outlier number is 1.71 per RX position, which is further processed by the subsequent out-of-Mahalanobis KPM re-clustering stage rather than being treated directly as a tracking failure.

For cluster tracking, the weights of centroid, shape, and density features in the MFM method are optimized according to the GCR criterion. Under this optimized weighting setting, the MFM algorithm achieves a GCR of 0.561. In comparison, the simultaneous clustering-and-tracking approach of MD-SCT reduces fragmentation and maintains cluster consistency over longer spatial ranges, as reflected by the lower GCR of 0.028. The MSSD, i.e., the mean square difference between adjacent values in a sequence, is computed for DB, CH, and the standard deviations of intra-cluster delay and angular parameters, where the notation $s_x$ denotes the MSSD of the variable $x$. The lower MSSD values of the MD-SCT algorithm indicate smaller variations between adjacent snapshots. Together with the lower GCR, the results show that the MD-SCT algorithm better maintains cluster-level spatial consistency across successive snapshots.

	\section{Conclusion}\label{section5}
	In this paper, we proposed a Mahalanobis-distance-based simultaneous clustering and tracking (MD-SCT) algorithm to capture the spatially consistent MPC evolution in cluster-based channel analysis. By computing the Mahalanobis distance over feature vectors constructed from parameters such as delay, angle, and transceiver location, the proposed technique performs clustering and tracking simultaneously without separating them into two independent stages. The algorithm is applied at 132 GHz in an industrial Internet of Things scenario. The evaluation based on MSSD and GCR demonstrates that the proposed algorithm improves the spatial consistency of clustered MPC sequences and cluster-center trajectories. These findings support spatial-consistency-aware channel modeling for 6G and can facilitate reliable beam tracking, massive MIMO, and integrated sensing and communication, especially in the sub-THz band.

   

	\bibliographystyle{IEEEtran}
	\bibliography{ref}

\end{document}